\documentclass[11pt]{article}
\usepackage{latexsym,epic,eepic}
\usepackage{amssymb}
\usepackage{geometry}
\usepackage{amsmath}
\usepackage{amsfonts}
\usepackage{graphicx}

\begin{document}
\newcommand{\2}{\vspace{0.2 cm}}
\newcommand{\dist}{{\rm dist}}
\newcommand{\diam}{{\rm diam}}
\newcommand{\rad}{{\rm rad}}
\newcommand{\dom}{\mbox{$\rightarrow$}}
\newcommand{\ldom}{\mbox{$\leftarrow$}}
\newcommand{\edom}{\mbox{$\leftrightarrow$}}
\newcommand{\ndom}{\mbox{$\not\rightarrow$}}
\newcommand{\sdom}{\mbox{$\Rightarrow$}}
\newcommand{\nsdom}{\mbox{$\not\Rightarrow$}}
\newcommand{\qed}{\hfill$\diamond$}
\newcommand{\pf}{{\bf Proof: }}
\newtheorem{theorem}{Theorem}[section]
\newcommand{\ra}{\rangle}
\newcommand{\la}{\langle}
\newtheorem{lemma}[theorem]{Lemma}
\newtheorem{corollary}[theorem]{Corollary}
\newtheorem{proposition}[theorem]{Proposition}
\newtheorem{conjecture}[theorem]{Conjecture}
\newtheorem{problem}[theorem]{Problem}
\newtheorem{remark}[theorem]{Remark}
\newtheorem{example}[theorem]{Example}
\newcommand{\beq}{\begin{equation}}
\newcommand{\eeq}{\end{equation}}
\newcommand{\argmax}{{\rm argmax}}
\newcommand{\MiP}{MinHOMP($H$) }
\newcommand{\MaP}{MaxHOMP($H$) }
\newcommand{\vecc}[1]{\stackrel{\leftrightarrow}{#1}}

\title{Complexity of the Minimum Cost Homomorphism Problem for
Semicomplete Digraphs with Possible Loops}
\author{Eun Jung Kim\thanks{Department of Industrial
Engineering, KAIST 373-1 Kusong-dong, Yusong-ku, Taejon, 305-791,
Republic of Korea, masquenada@kaist.ac.kr} \and Gregory Gutin\thanks
{Corresponding author. Department of Computer Science, Royal
Holloway University of London, Egham, Surrey TW20 OEX, UK,
gutin@cs.rhul.ac.uk and Department of Computer Science, University
of Haifa, Israel}}

\date{}

\maketitle

\begin{abstract}
For digraphs $D$ and $H$, a mapping $f:\ V(D)\dom V(H)$ is a
homomorphism of $D$ to $H$ if $uv\in A(D)$ implies $f(u)f(v)\in
A(H).$ For a fixed digraph $H$, the homomorphism problem is to
decide whether an input digraph $D$ admits a homomorphism to $H$ or
not, and is denoted as HOM($H$).

An optimization version of the homomorphism problem was motivated by
a real-world problem in defence logistics and was introduced in
\cite{gutinDAM154a}. If each vertex $u \in V(D)$ is associated with
costs $c_i(u), i \in V(H)$, then the cost of the homomorphism $f$ is
$\sum_{u\in V(D)}c_{f(u)}(u)$. For each fixed digraph $H$, we have
the {\em minimum cost homomorphism problem for} $H$ and denote it as
MinHOM($H$). The problem is to decide, for an input graph $D$ with
costs $c_i(u),$ $u \in V(D), i\in V(H)$, whether there exists a
homomorphism of $D$ to $H$ and, if one exists, to find one of
minimum cost.

Although a complete dichotomy classification of the complexity of
MinHOM($H$) for a digraph $H$ remains an unsolved problem, complete
dichotomy classifications for MinHOM($H$) were proved when $H$ is a
semicomplete digraph \cite{gutinDAM154b}, and a semicomplete
multipartite digraph \cite{gutinSIDMA, gutinDAM}. In these studies,
it is assumed that the digraph $H$ is loopless. In this paper, we
present a full dichotomy classification for semicomplete digraphs
with possible loops, which solves a problem in
\cite{gutinRMS}.\footnote{This paper was submitted to SIAM J.
Discrete Math. on October 27, 2006}
\end{abstract}

\section{Introduction, Terminology and Notation}\label{introsec}

For directed (undirected) graphs $G$ and $H$, a mapping $f:\
V(G)\dom V(H)$ is a {\em homomorphism of $G$ to $H$} if $uv$ is an
arc (edge) implies that $f(u)f(v)$ is an arc (edge). A homomorphism
$f$ of $G$ to $H$ is also called an {\em $H$-coloring} of $G$, and
$f(x)$ is called the {\em color} of the vertex $x$ in $G$. We denote
the set of all homomorphisms from $G$ to $H$ by $HOM(G,H)$.

Let $H$ be a fixed directed or undirected graph. The {\em
homomorphism problem}, HOM($H$), for $H$ asks whether a directed or
undirected input graph $G$ admits a homomorphism to $H.$ The {\em
list homomorphism problem}, ListHOM($H$), for $H$ asks whether a
directed or undirected input graph $G$ with lists (sets) $L_u
\subseteq V(H)$, admits a homomorphism $f$ to $H$ in which $f(u) \in
L_u$ for each $u \in V(G)$.

Suppose $G$ and $H$ are directed (or undirected) graphs, and
$c_i(u)$, $u\in V(G)$, $i\in V(H)$ are nonnegative {\em costs}. The
{\em cost of a homomorphism} $f$ of $G$ to $H$ is $\sum_{u\in
V(G)}c_{f(u)}(u)$. If $H$ is fixed, the {\em minimum cost
homomorphism problem}, MinHOM($H$), for $H$ is the following
optimization problem. Given an input graph $G$, together with costs
$c_i(u)$, $u\in V(G)$, $i\in V(H)$, find a minimum cost homomorphism
of $G$ to $H$, or state that none exists.

The minimum cost homomorphism problem was introduced in
\cite{gutinDAM154a}, where it was motivated by a real-world problem
in defence logistics. We believe it offers a practical and natural
model for optimization of weighted homomorphisms. The problem's
special cases include the list homomorphism problem
\cite{hell2003,hell2004} and the general optimum cost chromatic
partition problem, which has been intensively studied
\cite{halld2001,jansenJA34,jiangGT32}, and has a number of
applications \cite{kroon1997,supowitCAD6}.

If a directed (undirected) graph $G$ has no loops, we call $G$ {\em
loopless}. If a directed (undirected) graph $G$ has a loop at every
vertex, we call $G$ {\em reflexive}. When we wish to stress that a
family of digraphs may contain digraphs with loops, we will speak of
digraphs {\em with possible loops (w.p.l.)} For an undirected graph
$H$, $V(H)$ and $E(H)$ denote its vertex and edge sets,
respectively. For a digraph $H$, $V(H)$ and $A(H)$ denote its vertex
and arc sets, respectively.

In this paper, we give a complete dichotomy classification of the
complexity of MinHOM($H$) when $H$ is a semicomplete digraph with
possible loops. A dichotomy of MinHOM($H$) when $H$ is a tournament
w.p.l. was established in \cite{gutinRMS}, but it is much easier
than the more general dichotomy obtained in this paper. A full
dichotomy of MinHOM($H$) for $H$ being a (general) digraph has not
been settled yet and is considered to be a very difficult open
problem. Nonetheless, dichotomy have been obtained for special
classes of digraphs such as semicomplete digraphs and semicomplete
multipartite digraphs; see \cite{gutinDAM154b,gutinDAM, gutinSIDMA}.
Note that, for a semicomplete digraph $H$, the dichotomy for
MinHOM($H$) \cite{gutinDAM154b} is different from the dichotomy for
HOM($H$) \cite{bangSIAMJDM1} and ListHOM($H$) \cite{gutinDAM154b}
(the last two coincide). Our dichotomy for MinHOM($H$) when $H$ is a
semicomplete digraph w.p.l. is significantly more involved than the
one for MinHOM($H$) when $H$ is a (loopless) semicomplete digraph.

Notice that, apart from \cite{gutinRMS}, all these studies deal only
with loopless digraphs. When we study the structure of a digraph
$D$, we usually assume that $D$ has no loops. This is often a
natural assumption since many properties of loopless digraphs can
readily be extended to general digraphs w.p.l. as the loops do not
affect important parts of the structure of a digraph in the majority
of cases. When we investigate homomorphisms of undirected/directed
graphs, the situation is different and loops have to be taken into
consideration in the general case.

The homomorphism problem HOM($H$) is trivially polynomial time
solvable when $H$ has a loop, since we may simply map all the
vertices of the input graph to a vertex with a loop. However, if we
wish to get a dichotomy of MinHOM($H$) or ListHOM($H$), it is not
that simple. For example, in \cite{gutinEJC}, it turns out that the
class of proper interval graphs is exactly the class of reflexive
graphs for which MinHOM($H$) is polynomial time solvable (assuming,
as usual, that $P\neq NP$). On the other hand, if we assume that $H$
is loopless, MinHOM($H$) is polynomial time solvable if and only if
$H$ is a proper interval bigraph. It is often the case that, even if
we succeeded in obtaining a dichotomy classification of MinHOM($H$)
for reflexive and loopless $H$ separately, it is another issue to
get a dichotomy classification for $H$ with possible loops.

Complete dichotomy classifications of ListHOM($H$) and MinHOM($H$),
for an undirected graph $H$ w.p.l., have been achieved, see
\cite{federJCT, federCOM, federJGT} and \cite{gutinEJC}. For a
directed graph with possible loops, the study has just begun and
there are only a few results proved \cite{gutinRMS} so far. In
\cite{federMAN}, the authors prove some partial results on
complexity of ListHOM($H$) when $H$ is a reflexive digraph. In
particular, it is conjectured that for a reflexive digraph $H$,
ListHOM($H$) is polynomial time solvable if and only if $H$ has a
proper ordering. Here, we say that a reflexive digraph $H$ has a
{\em proper ordering} if its vertices can be ordered so that
whenever $xy,x'y'\in A(H)$, $\min({x,x'})\min({y,y'})$ is also in
$A(H)$. Unfortunately, the conjecture remains unconfirmed even for
the case of reflexive semicomplete digraphs.

In the rest of this section, we give additional terminology and
notation. In the subsequent sections, we first prove a full
dichotomy classification of the complexity of MinHOM($H$) when $H$
is a reflexive semicomplete digraph. Using this result, we will
further present a full dichotomy classification of MinHOM($H$) when
$H$ is a semicomplete digraph with possible loops.

For a digraph $D$, if $xy \in A(D)$, we say that $x$ {\em dominates}
$y$ and $y$ {\em is dominated} by $x$, denoted by $x \rightarrow y$.
Furthermore, if $xy \in A(D)$ and $yx\notin A(D)$, then we say that
$x$ {\em strictly dominates} $y$ and $y$ {\em is strictly dominated}
by $x$, denoted by $x\mapsto y$. For sets $X, Y \subseteq V(G)$, $X
\rightarrow Y$ means that $x \rightarrow y$ for each $x \in X$, $y
\in Y$. Also, for sets $X, Y \subseteq V(G)$, $X \mapsto Y$ means
that $xy\in A(D)$ but $yx \notin A(D)$ for each $x\in X$, $y\in Y$.
For $xy \in A(D)$, we call $xy$ an {\em asymmetric arc} if $yx\notin
A(D)$, and a {\em symmetric arc} if $yx \in A(D)$. A digraph $D$ is
{\em symmetric} if each arc of $D$ is symmetric. For a digraph $H$,
$H^{sym}$ denotes the {\em symmetric subdigraph} of $H$, i.e., a
digraph with $V(H^{sym})=V(H)$ and $A(H^{sym})=\{uv,vu \in A(H)\}$.
Note that any vertex $u$ of $V(H^{sym})$ has a loop if and only if
$u$ has a loop in $H$. We call a directed graph $D$ an {\em oriented
graph} if all arcs of $D$ are asymmetric.

For a digraph $D$, let $D[X]$ denote a subdigraph induced by $X
\subseteq V(D)$. For any pair of vertices of a directed graph $D$,
we say that $u$ and $v$ are {\em adjacent} if $u\rightarrow v$ or $v
\rightarrow u$, or both. The {\em underlying graph} $U(D)$ of a
directed graph $D$ is the undirected graph obtained from $D$ by
disregarding all orientations and deleting one edge in each pair of
parallel edges.  A directed graph $D$ is {\em connected} if $U(D)$
is connected. The {\em components} of $D$ are the subdigraphs of $D$
induced by the vertices of components of $U(D)$.

By a {\em directed path (cycle)} we mean a simple directed path
(cycle) (i.e., with no self-crossing). We assume that a directed
cycle has at least two vertices. In particular, a loop is not a
cycle. A directed cycle with $k$ vertices is called a {\em directed
$k$-cycle} and denoted by $\vec{C}_k.$ Let $K_n^*$ denote a complete
digraph with a loop at each vertex, i.e., a reflexive complete
digraph.

An {\em empty digraph} is a digraph with no arcs. A loopless digraph
$D$ is a {\em tournament} ({\em semicomplete digraph}) if there is
exactly one arc (at least one arc) between every pair of vertices.
We will consider {\em semicomplete digraphs with possible loops
(w.p.l.)}, i.e., digraphs obtained from semicomplete digraphs by
appending some number of loops (possibly zero loops). A {\em
$k$-partite tournaments} ({\em semicomplete $k$-partite digraph}) is
a digraph obtained from a complete $k$-partite graph by replacing
every edge $xy$ with one of the two arcs $xy,yx$ (with at least one
of the arcs $xy,yx$). An acyclic tournament on $p$ vertices is
denoted by $TT_p$ and called a {\em transitive tournament}. The
vertices of a transitive tournament $TT_p$ can be labeled
$1,2,\ldots ,p$ such that $ij\in A(TT_p)$ if and only if $1\le
i<j\le p.$ By $TT^-_p$ $(p\ge 2$), we denote $TT_p$ without the arc
$1p.$ For an acyclic digraph $H$, an ordering $u_1,u_2,\ldots ,u_p$
is called {\em acyclic} if $u_i\dom u_j$ implies $i<j.$

Let $H$ be a digraph. The {\em converse} of $H$ is the digraph
obtained from $H$ by replacing every arc $xy$ with the arc $yx.$ For
a pair $X,Y$ of vertex sets of a digraph $H$, we define $X\times
Y=\{xy:\ x\in X,y\in Y\}.$ Let $H$ be a loopless digraph with
vertices $x_1,x_2,\ldots ,x_p$ and let $S_1,S_2,\ldots ,S_p$ be
digraphs. Then the {\em composition} $H[S_1,S_2,\ldots ,S_p]$ is the
digraph obtained from $H$ by replacing $x_i$ with $S_i$ for each
$i=1,2,\ldots ,p.$ In other words, $$V(H[S_1,S_2,\ldots
,S_p])=V(S_1)\cup V(S_2)\cup \ldots \cup V(S_p) \mbox{ and}$$
$$A(H[S_1,S_2,\ldots ,S_p])=\cup\{V(S_i)\times V(S_j):\ x_ix_j\in A(H),\ 1\le i\neq
j\le p\}\cup (\cup_{i=1}^pA(S_i)).$$ If every $S_i$ is an empty
digraph, the composition $H[S_1,S_2,\ldots ,S_p]$ is called an {\em
extension} of $H.$

The {\em intersection graph} of a family $F=\{S_1,S_2, \ldots
,S_n\}$ of sets is the graph $G$ with $V(G)=F$ in which $S_i$ and
$S_j$ are adjacent if and only if $S_i\cap S_j\neq \emptyset$. Note
that by this definition, each intersection graph is reflexive. A
graph isomorphic to the intersection graph of a family of intervals
on the real line is called an {\em interval graph}. If the intervals
can be chosen to be inclusion-free, the graph is called a {\em
proper interval graph.}

\section{Classification for Reflexive Semicomplete Digraphs}\label{crsd}

In this section, we describe a dichotomy classification of the
complexity of MinHOM($H$) when $H$ is a reflexive semicomplete
digraph. Let $R$ be a reflexive digraph with $V(R)=\{1,2,3\}$ and
$A(R)=\{12,23,31,13,11,22,33\}$. Let $\vec{C}^*_3$ denote a
reflexive directed cycle on three vertices. The main dichotomy
classification of this section is given in the following theorem.

\begin{theorem}\label{refsd}
Let $H$ be a reflexive semicomplete digraph. If $H$ does not contain
either $R$ or $\vec{C}^*_3$ as an induced subdigraph, and
$U(H^{sym})$ is a proper interval graph (possibly with more than one
component), then MinHOM($H$) is polynomial time solvable. Otherwise,
MinHOM($H$) is NP-hard.
\end{theorem}

\subsection{NP-hard cases of MinHOM($H$)}
The following lemma is an obvious basic observation often used to
obtain dichotomies. This lemma is certainly applicable for a digraph
$H$ w.p.l.

\begin{lemma}\cite{gutinDAM154b}\label{reduction}
Let $H'$ be an induced subdigraph of a digraph $H$. If MinHOMP($H'$)
is NP-hard, then MinHOMP($H$) is also NP-hard.
\end{lemma}

The following assertion was proved in \cite{gutinRMS}.

\begin{lemma}\label{3cycle}
Let a digraph $H$ be obtained from $\vec{C}_k$, $k\ge 3$, by adding
at least one loop. Then MinHOM($H$) is NP-hard.
\end{lemma}

The following lemma shows that for a digraph $H$ obtained from
$\vec{C}_3$ by adding some loops and backward arcs, i.e., arcs of
the form $(i,i-1)$, MinHOM($H$) is NP-hard.

\begin{lemma}\label{3cyclesym}
Let $H$ be a digraph with $V(H)=\{1,2,3\}$ and
$A(H)=\{12,23,32,31,22,33\}\cup B$, where $B \subseteq \{11\}$. Then
MinHOM($H$) is NP-hard.
\end{lemma}
\pf Let $G$ be a loopless digraph with $p$ vertices. Construct a
bipartite digraph $D$ as follows: $V(D)=\{x_1,x_2:\ x\in V(G)\}$ and
$A(D)=\{x_1x_2:\ x\in V(G)\}\cup \{x_2y_1:\ xy\in A(G)\}.$ Set
$c_1(x_1)=0,$ $c_2(x_2)=3$, $c_2(x_1)=c_1(x_2)=4p+1$ and
$c_3(x_1)=c_3(x_2)=2$ for each $x\in V(G)$.

Clearly, $h(x_1)=h(x_2)=3$ for each $x\in V(D)$ defines a
homomorphism $h$ of $D$ to $H.$ Let $f$ be a minimum cost
homomorphism of $D$ to $H.$ It follows from the fact that the cost
of $h$ is $4p$ that $f(x_2)\neq 1$ and $f(x_1)\neq 2$ for each $x\in
V(G).$ Thus, for every arc $x_1x_2$ of $D$ we have three
possibilities of coloring: (a) $f(x_1)=1,f(x_2)=2$; (b)
$f(x_1)=f(x_2)=3$; (c) $f(x_1)=3,f(x_2)=2.$ Because of the three
choices and the structure of $H$, if $f(x_1)=3$ and $f(x_2)=2$, we
can recolor $x_2$ so that $f(x_2)=3$, decreasing the cost of $f$, a
contradiction. Thus, (c) is impossible for $f$.

Let $f(x_1)=f(y_1)=1$, where $x,y$ are distinct vertices of $G.$ If
$xy\in A(G)$, then $x_2y_1\in A(D)$, which is a contradiction since
$f(x_2)=2.$ Thus, $x$ and $y$ are non-adjacent in $G$. Hence,
$I=\{x\in V(G): f(x_1)=1\}$ is an independent set in $G$. Observe
that the cost of $f$ is $4p-|I|$.

Conversely, if $I$ is an independent set in $G$, we obtain a
homomorphism $g$ of $D$ to $H$ by fixing $g(x_1)=1$, $g(x_2)=2$ for
$x\in I$ and $g(x_1)=g(x_2)=3$ for $x\in V(G)-I$. Observe that the
cost of $g$ is $4p-|I|.$ Hence a homomorphism $g$ of $D$ to $H$ is
of minimum cost if and only if the corresponding independent set $I$
is of maximum size in $G$. Since the maximum size independent set
problem is NP-hard, MinHOM($H$) is NP-hard as well. Observe that the
validity of the proof does not depend on whether vertex 1 has a loop
or not. \qed

\begin{corollary}\label{refsd3cycle}
Let $H$ be a reflexive semicomplete digraph. If $H$ contains either
$R$ or $\vec{C}^*_3$ as an induced subdigraph, MinHOM($H$) is
NP-hard.
\end{corollary}
\pf It is straightforward to see that Lemmas \ref{3cycle} and
\ref{3cyclesym} imply the NP-hardness of MinHOM($\vec{C}^*_3$) and
MinHOM($R$), respectively. The above statement follows directly from
Lemma \ref{reduction}. \qed

The following theorem is from \cite{gutinEJC}.

\begin{theorem}\label{refundich}
Let $H$ be a reflexive graph. If $H$ is a proper interval graph
(possibly with more than one component), then the problem
MinHOM($H$) is polynomial time solvable. In all other cases, the
problem MinHOM($H$) is NP-hard.
\end{theorem}

Suppose that $H$ is a semicomplete digraph and that $U(H^{sym})$ is
not a proper interval graph. That is, at least one component of
$U(H^{sym})$ is not a proper interval graph. Then
MinHOM($U(H^{sym})$) is polynomial time reducible to MinHOM($H$)
since an input graph $G$ of MinHOM($U(H^{sym})$) can be transfomed
into an input digraph $G^*$ of MinHOM($H$) by replacing each edge
$xy$ of $G$ by a symmetric arc $xy$ of $G^*.$ Hence, if $U(H^{sym})$
is not a proper interval graph, MinHOM($H$) is NP-hard by Theorem
\ref{refundich}. Together with Corollary \ref{refsd3cycle}, this
proves the claim for the NP-hardness part of Theorem \ref{refsd}.

\begin{theorem}\label{refsdnh}
Let $H$ be a reflexive semicomplete digraph. If $H$ contains either
$R$ or $\vec{C}^*_3$ as an induced subdigraph, or  $U(H^{sym})$ is
not a proper interval graph, then MinHOM($H$) is NP-hard.
\end{theorem}

\subsection{Polynomial time solvable cases of MinHOM($H$)}

Let $H$ be a digraph and let $v_1,v_2, \ldots, v_p$ be an ordering
of $V(H).$ Let $e=v_iv_r$ and $f=v_jv_s$ be two arcs in $H$. The
pair $v_{\min\{i,j\}}v_{\min\{s,r\}}$
($v_{\max\{i,j\}}v_{\max\{s,r\}}$) is called the {\em minimum} ({\em
maximum}) of the pair $e,f$. (The minimum (maximum) of two arcs is
not necessarily an arc.) An ordering $v_1,v_2, \ldots, v_p$ is a
{\em Min-Max ordering of $V(H)$} if both minimum and maximum of
every two arcs in $H$ are in $A(H).$ Two arcs $e,f\in A(H)$ are
called a {\em non-trivial pair} if $\{e,f\}\neq \{g',g''\}$, where
$g'$ ($g''$) is the minimum (maximum) of $e,f.$ Clearly, to check
that an ordering is Min-Max, it suffices to verify that the minimum
and maximum of every non-trivial pair of arcs are arcs, too.

The following theorem was proved in \cite{gutinDAM154b} for loopless
digraphs. In fact, the same proof is valid for digraphs with
possible loops.

\begin{theorem}\label{mmth}
Let $H$ be a digraph and let an ordering $1,2,\ldots, p$ of $V(H)$
be a Min-Max ordering, i.e., for any pair $ik,js$ of arcs in $H$, we
have $\min\{i,j\}\min\{k,s\}\in A(H)$ and $\max\{i,j\}\max\{k,s\}\in
A(H).$ Then MinHOM($H$) is polynomial time solvable.
\end{theorem}

In this subsection, we assume that $H$ is a reflexive semicomplete
digraph which contains neither $R$ nor $\vec{C}^*_3$, and for which
$U(H^{sym})$ is a proper interval graph (possibly with more than one
component), unless we mention otherwise. In this subsection, we will
show that $H$ has a Min-Max ordering, and, thus, MinHOM($H$) is
polynomial time solvable by Theorem \ref{mmth}.



There is a useful characterization of proper interval graphs
\cite{hellSIDMA18, spinrad2003}.

\begin{theorem}\label{pincha}
A reflexive graph $H$ is a proper interval graph if and only if its
vertices can be ordered $v_1,v_2,\ldots ,v_n$ so that $i<j<k$ and
$v_iv_k\in E(H)$ imply that $v_iv_j\in E(H)$ and $v_jv_k \in E(H)$.
\end{theorem}

Let $H$ be a digraph and let $v_1,v_2, \ldots, v_p$ be an ordering
of $V(H).$ We call $v_iv_j$ a {\em forward arc} (with respect to the
ordering) if $i<j$, and a {\em backward arc} if $i>j$. The following
lemma shows that if $H$ satisfies a certain condition, then the
vertices of $H$ can ordered so that every arc is either forward or
symmetric.

\begin{lemma}\label{pinord}
Let $H$ be a reflexive semicomplete digraph and suppose $H$ does not
contain $R$ as an induced subdigraph and suppose that $U(H^{sym})$
is a connected proper interval graph. Then the vertices of $H$ can
be ordered $v_1,v_2,\ldots ,v_n$ such that $i<j<k$ and $v_iv_k\in
A(H^{sym})$ imply that $v_iv_j\in A(H^{sym})$ and $v_jv_k \in
A(H^{sym})$ and furthermore, for every pair of vertices $v_i$ and
$v_j$ with $i<j$, we have $v_i\rightarrow v_j$.
\end{lemma}
\pf Since $U(H^{sym})$ is a proper interval graph, the vertices of
$H$ can be ordered $v_1,\ldots ,v_n$ such that $i < j < k$ and
$v_iv_k\in A(H^{sym})$ imply that $v_iv_j\in A(H^{sym})$ and $v_jv_k
\in A(H^{sym})$ by Theorem \ref{pincha}. Observe that if $v_iv_j$ is
a symmetric arc with $i<j$, then for each $\ell,k$ with $i<\ell<k<j$
we have $v_{\ell}v_k$ is a symmetric arc. Note also that
$v_iv_{i+1}$ for each $i=1,\ldots ,n-1$ is a symmetric arc, since
otherwise $H^{sym}$ has more that one component, contradicting the
connectivity assumption.

We wish to prove that if $v_{\ell}\rightarrow v_k$ for some
$\ell<k$, then $v_i\rightarrow v_j$ for each $i<j.$ We prove it
using a sequence of claims.

{\bf Claim 1.} If $v_i \mapsto v_j$ for some $j>i$, then $v_i
\rightarrow \{v_{i+1},\ldots ,v_n\}$.

\noindent{\bf Proof:} If $v_iv_k\in A(H^{sym})$ for each $k>i$,
there is nothing to prove. Thus, we may assume without loss of
generality that there exists a vertex $v_k$ such that $v_k \mapsto
v_i$. By an observation above, all arcs between $v_i$ and $v_t$ for
each $t\ge \min\{j,k\}$ are asymmetric. Thus, there is an index
$m\ge \min\{j,k\}$ such that either $v_i \mapsto v_m$ and $v_{m+1}
\mapsto v_i$ or $v_m \mapsto v_i$ and $v_i \mapsto v_{m+1}.$ Recall
that that $v_mv_{m+1}$ is a symmetric arc. Hence, $H[\{v_i, v_m,
v_{m+1}\}]\cong R$, a contradiction.

A similar argument leads to the symmetric statement below.

{\bf Claim 1$'$.} If $v_j \mapsto v_i$ for some $j<i$, then
$\{v_1,\ldots ,v_{i-1}\}\rightarrow v_i$.

{\bf Claim 2.} We have either $\{v_1,\ldots ,v_{i-1}\}\rightarrow
v_i \rightarrow \{v_{i+1},\ldots ,v_n\}$ or $\{v_{i+1},\ldots
,v_n\}\rightarrow v_i \rightarrow \{v_1,\ldots ,v_{i-1}\}$.

\noindent{\bf Proof:} Suppose to the contrary that there are two
vertices $v_j,v_k$ with $j<i<k$ such that $v_i \mapsto v_j$,
$v_i\mapsto v_k$ in $H$. (The case for which $v_j \mapsto v_i$,
$v_k\mapsto v_i$ in $H$ can be treated in a similar manner.) Then
$v_jv_k$ is not a symmetric arc since otherwise, $v_jv_i$ and
$v_iv_k$ must be symmetric arcs by the property of the ordering.
Hence, only one of $v_jv_k$ and $v_kv_j$ is an arc of $H$. In either
case, we have a contradiction by Claims 1 or 1$'$.

{\bf Claim 3:} If $v_{\ell}\rightarrow v_k$ for some $\ell<k$, then
$v_i\rightarrow v_j$ for each $i<j.$

\noindent{\bf Proof:} Suppose to the contrary that there exist two
vertices $v_{j}$ and $v_{i}$ such that $v_{j} \mapsto v_{i}$ and
$i<j$. If any two of the four vertices $v_i, v_j, v_{\ell}$ and
$v_{k}$ are identical, we have a contradiction by Claim 1, 1$'$ or
2. Thus, we may assume that these vertices are all distinct. We have
the following cases.

(a) Let $i < \ell$. Then we have $i<\ell<k$. If $v_{i}v_k$ is a
symmetric arc, the arc $v_{\ell}v_k$ must be symmetric by the
property of the ordering, a contradiction. Hence, only one of
$v_{i}v_k$ and $v_kv_{i}$ is an arc of $H$. If $v_i\mapsto v_k$, we
have a contradiction by Claim 1 for vertex $v_i$ and if $v_k\mapsto
v_i$, we have a contradiction by Claim 1$'$ for vertex $v_k$.

(b) Let $\ell < i$. Then we have $\ell < i<j$. If $v_{\ell}v_{j}$ is
a symmetric arc, then $v_{i}v_{j}$ must be a symmetric arc by the
property of the ordering, contradiction. Hence, only one of
$v_jv_{\ell}$ and $v_{\ell}v_j$ is an arc of $H$. In either case, we
have a contradiction by Claims 1 or 1$'$.

By Claim 3, either $v_1,v_2,\ldots ,v_n$ or its reversal satisfies
the required property. \qed

\2

Consider a reflexive semicomplete digraph $H$. Suppose that $H$ does
not contain either $R$ or $\vec{C}^*_3$ as an induced subdigraph and
$U(H^{sym})$ is a proper interval graph. Note that each isolated
vertex in $U(H^{sym})$ forms a trivial proper interval graph in
itself.

Suppose that $H^{sym}$ is not connected. If each component of
$H^{sym}$ is trivial, it is clear that $H$ has a Min-Max ordering
since $H$ is a reflexive transitive tournament ($H$ does not contain
$\vec{C}^*_3$). Hence we may assume that at least one component of
$H^{sym}$ is nontrivial. Let $H^{sym}_i$ and $H^{sym}_j$ be two
distinct components of $H^{sym}$ and at least one of them, say
$H^{sym}_j$, is a nontrivial component containing more than one
vertex. Clearly, the arcs between $H^{sym}_i$ and $H^{sym}_j$ are
all asymmetric.

Let $u$ be a vertex of $H^{sym}_i$, and let $v$ and $w$ be two
distinct vertices in $H^{sym}_j$. Without loss of generality, we may
assume that $u \mapsto v$. If $w \mapsto u$, there must exist
adjacent vertices $p$ and $q$ on a path from $v$ to $w$ in
$H^{sym}_j$ such that $u \mapsto p$ and $q \mapsto u$. Then we have
$H[\{u, p, q\}]\cong R$, a contradiction. With a similar argument,
it is easy to see that all arcs between two components $H^{sym}_i$
and $H^{sym}_j$ are oriented in the same direction with respect to
the components. Furthermore, since $H$ is $\vec{C}^*_3$-free, the
components of $H^{sym}$ can be ordered $H^{sym}_1, H^{sym}_2, \ldots
,H^{sym}_l$ so that for each pair of vertices $u \in H^{sym}_i$ and
$v \in H^{sym}_j$ with $i < j$, we have $u \mapsto v$. This implies
the following:

\begin{corollary}\label{compord} Let $H$ be a reflexive
semicomplete digraph and suppose $H$ does not contain either $R$ or
$\vec{C}^*_3$ as an induced subdigraph and suppose that $U(H^{sym})$
is a proper interval graph. Then the components of $H^{sym}$ can be
ordered $H^{sym}_1, H^{sym}_2, \ldots ,H^{sym}_l$ such that if $u
\in H^{sym}_i$, $v \in H^{sym}_j$ and $i < j$, then we have $u
\mapsto v$.
\end{corollary}

We shall call the ordering of the components of $H^{sym}$ described
in Corollary \ref{compord} an {\em acyclic ordering} of the the
components of $H^{sym}$. Now, with Lemma \ref{pinord}, we have the
following lemma.

\begin{lemma}\label{refsdord}
Let $H$ be a reflexive semicomplete digraph and suppose $H$ does not
contain either $R$ or $\vec{C}^*_3$ as an induced subdigraph and
$U(H^{sym})$ is a proper interval graph. Then the vertices of $H$
can be ordered $v_1, \ldots ,v_n$ so that $i<j<k$ and $v_iv_k\in
A(H^{sym})$ imply that $v_iv_j\in A(H^{sym})$ and $v_jv_k \in
A(H^{sym})$ and furthermore, for every pair of vertices $v_i$ and
$v_j$ from $V(H)$ with $i<j$, we have $v_i\rightarrow v_j$.
\end{lemma}
\pf Let $H^{sym}_1, H^{sym}_2, \ldots ,H^{sym}_l$ be the acyclic
ordering of the components of $H^{sym}$. By Lemma \ref{pinord}, we
have an ordering $v_1^i,v_2^i, \dots ,v_{|V(H^{sym}_i)|}^i$ of
$V(H^{sym}_i)$, for each $i=1,\ldots ,l$, such that every asymmetric
arc is forward. Then the ordering $$v_1^1,v_2^1, \ldots
,v_{|V(H^{sym}_1)|}^1,v_1^2,v_2^2, \ldots
,v_{|V(H^{sym}_2)|}^2,\ldots , v_1^l,v_2^l, \ldots
,v_{|V(H^{sym}_l)|}^l$$ of the vertices of $H$ satisfies the
condition, completing the proof. \qed

\2

The following proposition was proved in \cite{gutinEJC}. Observe
that for the symmetric subdigraph $H^{sym}$, the ordering of the
vertices of $H$ described in Lemma \ref{refsdord} satisfies the
condition of the proposition below.

\begin{proposition}\label{refmm}
A reflexive graph $H$ has a Min-Max ordering if and only if its
vertices can be ordered $v_1, v_2, \ldots ,v_n$ so that $i<j<k$ and
$v_iv_k\in E(H)$ imply that $v_iv_j\in E(H)$ and $v_jv_k \in E(H)$.
\end{proposition}

\begin{lemma}\label{refsdmm}
Let $H$ be a reflexive semicomplete digraph. If $H$ does not contain
either $R$ or $\vec{C}^*_3$ as an induced subdigraph, and
$U(H^{sym})$ is a proper interval graph, then $H$ has a Min-Max
ordering.
\end{lemma}

\pf Let $v_1,\ldots ,v_n$ be the ordering of $V(H)$ as described in
Lemma \ref{refsdord}. We will show that this is a Min-Max ordering
of $V(H)$.

Let $v_iv_j$ and $v_kv_l$ be a non-trivial pair of $H$. If $i\leq j$
and $k\leq l$, it is easy to see that both the minimum and maximum
of $v_iv_j$ and $v_kv_l$ are in $A(H)$. If $i>j$ and $k>l$, $v_iv_j$
and $v_kv_l$ are symmetric arcs of $H^{sym}$. Since they are a
non-trivial pair, the vertices $v_i,v_j,v_k$ and $v_l$ belong to the
same component of $H^{sym}$ by the proof of Lemma \ref{refsdord}.
Then the minimum and the maximum of $v_iv_j$ and $v_kv_l$ are also
in $A(H^{sym})$ by Lemma \ref{pinord}.

Now suppose that $i\leq j$ and $k>l$. Note that $v_kv_l$ is a
symmetric arc. Hence if $i=j$, then the vertices $v_i,v_k$ and $v_l$
belong to the same component of $H^{sym}$, and, thus, the minimum
and the maximum of $v_iv_j$ and $v_kv_l$ are in $A(H^{sym})$ by
Lemma \ref{pinord}. If $i\neq j$, we need to consider the following
four cases covering all possibilities for non-trivial pairs:

(a) $i \leq l< j\leq k$. Then $v_{\min\{i,k\}}v_{\min\{j,l\}}=v_iv_l
\in A(H)$ as $i\leq l$. Also, since $v_lv_k$ is a symmetric arc,
$v_{\max\{i,k\}}v_{\max\{j,l\}}=v_kv_j$ is a symmetric arc by Lemma
\ref{refsdord}.

(b) $l<i< j\leq k$. Then, since $v_lv_k$ is a symmetric arc,
$v_{\max\{i,k\}}v_{\max\{j,l\}}=v_kv_j$ and
$v_{\min\{i,k\}}v_{\min\{j,l\}}=v_iv_l$ are symmetric arcs by Lemma
\ref{refsdord}.

(c) $l<i < k <j$. Then, $v_{\max\{i,k\}}v_{\max\{j,l\}}=v_kv_j \in
A(H)$ as $k<j$. Also, since $v_lv_k$ is a symmetric arc,
$v_{\min\{i,k\}}v_{\min\{j,l\}}=v_iv_l$ is a symmetric arc by Lemma
\ref{refsdord}.

(d) $i\leq l< k <j$. Then $v_{\min\{i,k\}}v_{\min\{j,l\}}=v_iv_l \in
A(H)$ and $v_{\max\{i,k\}}v_{\max\{j,l\}}=v_kv_j \in A(H)$ as $i\le
l$ and $k< j$. \qed

\begin{theorem}\label{refsdpoly}
Let $H$ be a reflexive semicomplete digraph. If $H$ does not contain
either $R$ or $\vec{C}^*_3$ as an induced subdigraph, and
$U(H^{sym})$ is a proper interval graph, then MinHOM($H$) is
polynomial time solvable.
\end{theorem}
\pf This is a direct consequence of Lemmas \ref{refsdmm} and
\ref{mmth}. \qed

\begin{corollary}
Suppose $P\neq NP$. Let $H$ be a reflexive semicomplete digraph.
Then MinHOM($H$) is polynomial time solvable if and only if $H$ has
a Min-Max ordering.
\end{corollary}
\pf This is a direct consequence of Lemma \ref{refsdmm} and Theorem
\ref{refsdnh}.

\section{Classification for Semicomplete Digraphs with Possible Loops}\label{csd}

In this section, we describe a dichotomy classification for
MinHOM($H$) when $H$ is a semicomplete digraph with possible loops.
Let $W$ be a digraph with $V(W)=\{1,2\}$ and $A(W)=\{12,21,22\}$.
Let $R'$ be a digraph with $V(R')=\{1,2,3\}$ and
$A(R')=\{12,23,32,31,22,33\}$.

Given a semicomplete digraph $H$ w.p.l., let $L=L(H)$ and $I=I(H)$
denote the maximal induced subdigraphs of $H$ which are reflexive
and loopless, respectively. When $H=L$, we have obtained a dichotomy
classification for reflexive semicomplete digraph in Section
\ref{crsd}. When $H=I$, we also have a dichotomy classification by
the following theorem from \cite{gutinDAM154b}.

\begin{theorem}\label{sddich}
For a semicomplete digraph $H$, MinHOM($H$) is polynomial time
solvable if $H$ is acyclic or $H=\vec{C}_k$ for $k=$2 or 3, and
NP-hard, otherwise.
\end{theorem}

In this section, we will show that the following dichotomy
classification holds when $H$ is a semicomplete digraph w.p.l.

\begin{theorem}\label{wplsd}
Let $H$ be a semicomplete digraph with possible loops. If one of the
following holds, then MinHOM($H$) is polynomial time solvable.
Otherwise, it is NP-hard.

(i) The digraph $H=\vec{C}_k$ for $k=$2 or 3.

(ii-a) The digraph $L$ does not contains either $R$ or $\vec{C}^*_3$
as an induced subdigraph, and $U(L^{sym})$ is a proper interval
graph; $I$ is a transitive tournament; $H$ does not contain either
$W$, $R'$ or $\vec{C}_3$ with at least one loop as an induced
subdigraph.

or equivalently,

(ii-b) The digraph $H=TT_{k}[S_1,S_2,\ldots ,S_{k}]$ where $S_i$ for
each $i=1, \ldots ,k$ is either a single vertex without a loop, or a
reflexive semicomplete digraph which does not contain $R$ as an
induced subdigraph and for which $U(S^{sym}_i)$ is a connected
proper interval graph.
\end{theorem}

Through subsections \ref{wplnpcase} and \ref{wplpolycase}, we will
consider only the polynomiality condition (ii-a) in Theorem
\ref{wplsd}. We first prove the NP-hardness part of Theorem
\ref{wplsd} in subsection \ref{wplnpcase}. In subsection
\ref{wplpolycase}, a proof for the polynomial solvable case is
given. Finally in subsection \ref{wplchar}, we will prove the
equivalence of condition (ii-a) and (ii-b) in Theorem \ref{wplsd}.

\subsection{NP-hard cases of MinHOM($H$)}\label{wplnpcase}

The following two lemmas were proved in \cite{gutinRMS}.

\begin{lemma}\label{2cycle}
MinHOM($W$) is NP-hard.
\end{lemma}

\begin{lemma}\label{domin}
Let $H'$ be a digraph obtained from $\vec{C}_k=12\ldots k1$, $k\ge
2$, by adding an extra vertex $k+1$ dominated by at least two
vertices of the cycle and let $H''$ is the digraph obtained from
$H'$ by adding the loop at vertex $k+1.$ Let $H$ be $H'$ or its
converse or $H''$ or its converse. Then MinHOM($H$) is NP-hard.
\end{lemma}

Observe that MinHOM($R'$) is NP-hard by Lemma \ref{3cyclesym}. The
following result was proved in \cite{bangSIAMJDM1}.

\begin{theorem}\label{2cycleth}
Let $H$ be a (loopless) semicomplete digraph with at least two
directed cycles. Then the problem of checking whether a digraph $D$
has an $H$-coloring is NP-complete.
\end{theorem}

\begin{lemma}\label{3cyclesym2}
Let $H$ be a digraph with $V(H)=\{1,2,3\}$ and
$A(H)=\{12,21,23,31,33\}$. Then MinHOM($H$) is NP-hard.
\end{lemma}

\pf We will reduce the maximum independent set problem to
MinHOM($H$). Before we do this we consider a digraph $D^{*}(u,v)$
defined as follows. Here we set $e=uv$:
\begin{displaymath}
V(D^{*}(u,v))=\{u^e,u,v^e,v,x_1^e,x_2^e,\ldots ,x_6^e\}
\end{displaymath}
\begin{displaymath}
A(D^{*}(u,v))=\{x_1^ex_2^e,x_2^ex_3^e,\ldots
,x_5^ex_6^e,x_6^ex_1^e,x_4^eu^e,u^eu,x_5^ev^e,v^ev\}
\end{displaymath}

Let $G$ be a graph with $p$ vertices. Construct a digraph $D$ as
follows: Start with $V(D)=V(G)$ and, for each edge $e=uv\in E(G)$,
add a distinct copy of $D^{*}(u,v)$ to $D$. Note that the vertices
in $V(G)$ form an independent set in $D$ and that
$|V(D)|=|V(G)|+8|E(G)|$.

Given an edge $e=uv\in E(G)$, we fix the costs as follows: Let
$c_1(x_1^e)=0$ and $c_i(x_1^e)=p+1$ for each $i=2,3$. Let
$c_i(x_j^e)=0$ for each $i=1,2,3$ and $j=2,\ldots ,6$ apart from
$c_3(x_4^e)=c_3(x_5^e)=p+1$. Also, $c_i(u^e)=c_i(v^e)=0$ for each
$i=1,2,3$, $c_2(u)=c_2(v)=0$, $c_1(u)=c_1(v)=1$ and
$c_3(u)=c_3(v)=p+1$.

Consider a mapping $h$ of $V(D)$ to $V(H)$ as follows: $h(x_i^e)=1$
if $i$ is odd, $h(x_i^e)=2$ if $i$ is even, $h(u^e)=3$, $h(v^e)=2$
for each $e\in E(G)$ and $h(u)=1$ for each $u\in V(G)$. It is easy
to check that $h$ defines a homomorphism of $D$ to $H$ and the cost
of $h$ is $p$. Let $f$ be a minimum cost homomorphism of $D$ to $H$.
It follows from the fact that the cost of $h$ is $p$ that
$f(x_1^e)=1$, $f(x_4^e),f(x_5^e)\in \{1,2\}$ for each $e\in E(G)$,
and $f(u)\in \{1,2\}$ for each each $u\in V(G)$. Moreover, due to
the structure of $D^{*}(u,v)$ and the costs, for each $e\in E(G)$,
$(f(x_1^e),\ldots ,f(x_6^e))$ must coincide with one of the
following two sequences: (1,2,1,2,1,2) or (1,2,3,1,2,3).

If the first sequence is the actual one, then we have $f(x_4^e)=2$,
$f(u^e)\in \{1,3\}$, $f(u)\in \{1,2\}$ and $f(x_5^e)=1$, $f(v^e)=2$,
$f(v)=1$. If the second sequence is the actual one, we have
$f(x_4^e)=1$, $f(u^e)=2$, $f(u)=1$ and $f(x_5^e)=2$, $f(v^e)\in
\{1,3\}$, $f(v)\in \{1,2\}$. So in both cases we can assign both $u$
and $v$ color 1. Furthermore, by choosing the right sequence we can
color one of $u$ and $v$ with color 2 and the other with color 1.
Notice that $f$ cannot assign color 2 to both $u$ and $v$.

Clearly, $f$ must assign as many vertices of $V(G)$ in $D$ color 2.
However, if $uv$ is an edge in $G$, by the argument above, $f$
cannot assign color 2 to both $u$ and $v$. Hence, $I=\{u\in V(G):\
f(u)=2\}$ is an independent set in $G$. Observe that the cost of $f$
is $p-|I|$.

Conversely, if $I'$ is an independent set in $G$, we obtain a
homomorphism $g$ of $D$ to $H$ by fixing $g(u)=2$ for $u\in I'$,
$g(u)=1$ for $u\notin I'$. We can choose an appropriate sequence for
$x_1^e,\ldots, x_6^e$ for each edge $e \in E(G)$ and fix the
assignment of $u^e$ and $v^e$ accordingly by the above argument.
Observe that the cost of $g$ is $p-|I'|.$ Hence the cost of a
minimum homomorphism $f$ of $D$ to $H$ is $p-\alpha$, where $\alpha$
is the size of the maximum independent set in $G$. Since the maximum
size independent set problem is NP-hard, MinHOM($H$) is NP-hard as
well.\qed

\begin{lemma}\label{3cyclesym3}
Let $H$ be a digraph with $V(H)=\{1,2,3\}$ and
$A(H)=\{12,21,23,31\}\cup B_1 \cup B_2$, where $B_1$ is either
$\{11,22\}$ or $\emptyset$ and $B_2$ is either $\{33\}$ or
$\emptyset$. Then MinHOM($H$) is NP-hard.
\end{lemma}
\pf Consider the following three cases.

{\bf Case 1:} $B_1=\emptyset$ and $B_2=\emptyset$. Then MinHOM($H$)
is NP-hard by Theorem \ref{2cycleth}.

{\bf Case 2:} $B_1=\emptyset$ and $B_2=\{33\}$. Then MinHOM($H$) is
NP-hard by Lemma \ref{3cyclesym2}

{\bf Case 3:} $B_1=\{11,22\}$. Then MinHOM($H$) is NP-hard by Lemma
\ref{3cyclesym}.\qed

\2

Consider a strong semicomplete digraph w.p.l. $H$ on three vertices.
We want to obtain all polynomial cases for $H$. If $H$ does not have
a 2-cycle, MinHOM($H$) is NP-hard if (and only if) at least one of
its vertices has a loop by Lemma \ref{3cycle}. Note that we have a
polynomial case if $H$ is $\vec{C}_3$. Suppose that $H$ has at least
one 2-cycle. If there are two or more 2-cycles, MinHOM($H$) is
NP-hard by Theorem \ref{2cycleth} and Lemma \ref{2cycle} unless $H$
is reflexive. Note that in the reflexive case, MinHOM($H$) is
polynomial time solvable since $H$ has a Min-Max ordering.

Now suppose that $H$ has only one 2-cycle. For MinHOM($H$) to be not
NP-hard, both or neither of the two vertices forming the 2-cycle
must have loops simultaneously since otherwise, MinHOM($H$) is
NP-hard by Lemma \ref{2cycle}. Now by Lemma \ref{3cyclesym3},
MinHOM($H$) is still NP-hard.

Let $K^*_3-e$ be a digraph obtained by removing a nonloop arc from
$K^*_3$. The above observation can be summarized by the following
statement.

\begin{corollary}\label{3cycledich}
Let $H$ be a strong semicomplete digraph w.p.l. on three vertices.
If $H$ is either $\vec{C}_3$, $K^*_3$ or $K^*_3-e$, MinHOM($H$) is
polynomial time solvable. Otherwise, MinHOM($H$) is NP-hard.
\end{corollary}

For a semicomplete digraph w.p.l. $H$, if either MinHOM($L$) or
MinHOM($I$) is NP-hard, MinHOM($H$) is NP-hard by Lemma
\ref{reduction}. (Recall that $L=L(H)$ and $I=I(H)$ denote the
maximal induced subdigraphs of $H$ which are reflexive and loopless,
respectively.) Also, if $H$ contains $\vec{C}_3$ with at least one
loop as an induced subdigraph, MinHOM($H$) is NP-hard by Lemmas
\ref{3cycle} and \ref{reduction}. Suppose MinHOM($H$) is not
NP-hard. Then Lemma \ref{2cycle} indicates that for any pair of
vertices $u\in V(L)$ and $v\in V(I)$, either $u \rightarrow v$ or $v
\rightarrow u$, not both, as otherwise MinHOM($H$) is NP-hard. With
these observations and Lemma \ref{domin}, the following statement is
easily derived.

\begin{lemma}\label{wplsdnhl}
Let $H$ be a semicomplete digraph. If one of the following condition
holds, MinHOM($H$) is NP-hard.

(a) $I$ contains a cycle and $I\neq \vec{C}_k$ for $k=$2 or 3.

(b) $L$ contains either $R$ or $\vec{C}^*_3$ as an induced
subdigraph, or $U(L^{sym})$ is not a proper interval graph.

(c) $I=\vec{C}_k$ for $k=$2 or 3, and $L$ is nonempty.

(d) $H$ contains $W$, $R'$ or $\vec{C}_3$ with at least one loop as
an induced subdigraph.
\end{lemma}
\pf If condition (a) holds, MinHOM($H$) is NP-hard by Theorem
\ref{sddich} and Lemma \ref{reduction}. If condition (b) holds,
MinHOM($H$) is NP-hard by Theorem \ref {refsd} and Lemma
\ref{reduction}. If condition (d) holds, MinHOM($H$) is NP-hard by
Lemmas \ref{2cycle}, \ref{3cyclesym}, \ref{3cycle} and
\ref{reduction}. The only remaining part is to prove that the
condition (c) is sufficient for MinHOM($H$) to be NP-hard.

If $I=\vec{C}_2$ and $u$ is a vertex of $L$, then we may assume that
either $u$ dominates both vertices of $I$, or $u$ is dominated by
one of $V(I)$ and dominates the other without loss of generality. In
the former case, MinHOM($H$) is NP-hard by Lemma \ref{domin}. In the
latter case, MinHOM($H$) is NP-hard by Lemma \ref{3cyclesym2}. If
$I=\vec{C}_3$ and $u$ is a vertex with a loop, MinHOM($H$) is
NP-hard by Lemma \ref{domin}. \qed

In fact, Lemma \ref{wplsdnhl} proves the NP-hardness part in Theorem
\ref{wplsd}. This can be seen as follows. Suppose that  MinHOM($H$)
is not NP-hard. Recall that the polynomiality conditions of Theorem
\ref{wplsd} are: (i) $H=\vec{C}_k$ for $k=$2 or 3, or (ii-a) $L$
does not contain either $R$ or $\vec{C}^*_3$ as an induced
subdigraph,  and $U(L^{sym})$ is a proper interval graph, $I$ is a
transitive tournament, and $H$ does not contain either $W$, $R'$ or
$\vec{C}_3$ with at least one loop as an induced subdigraph.

Suppose that a semicomplete digraph w.p.l. $H$ has an loopless
cycle. Then condition (ii-a) does not hold, and for condition (i) to
be violated, either one of (a) and (c) in Lemma \ref{wplsdnhl} must
hold. On the other hand, suppose that the loopless part $I$ of $H$
is a transitive tournament. Then condition (i) does not hold, and
for condition (ii-a) to be violated, one of (b) and (d) in Lemma
\ref{wplsdnhl} must hold.

\begin{corollary}\label{wplsdnh}
Let $H$ be a semicomplete digraph with possible loops. If none of
the following holds, then MinHOM($H$) is NP-hard.

(a) The digraph $H=\vec{C}_k$ for $k=$2 or 3.

(b) The digraph $L$ does not contains either $R$ or $\vec{C}^*_3$ as
an induced subdigraph, and $U(L^{sym})$ is a proper interval graph;
$I$ is a transitive tournament; $H$ does not contain either $W$,
$R'$ or $\vec{C}_3$ with at least one loop as an induced subdigraph.
\end{corollary}

\subsection{Polynomial time solvable cases of MinHOM($H$)}\label{wplpolycase}

If condition (i) in Theorem \ref{wplsd} holds for a semicomplete
digraph w.p.l. $H$, MinHOM($H$) is clearly polynomial time solvable
by Theorem \ref{sddich}. Although $\vec{C}_3$ does not have a
Min-Max ordering, there is a simple algorithm which solves
MinHOM($H$) in polynomial time when $H=\vec{C}_k$, $k \geq 2$, see
\cite{gutinDAM154b,gutinRMS}.

Therefore, we only need to prove that when $H$ satisfies the
condition (ii-a) in Theorem \ref{wplsd}, MinHOM($H$) is polynomial
time solvable. We claim that $H$ has a Min-Max ordering in this
case. Before showing this claim, we prove that the ordering
described in Lemma \ref{refsdord} for a reflexive semicomplete
digraph can be extended to a semicomplete digraph w.p.l. if
condition (ii-a) in Theorem \ref{wplsd} is satisfied.

\begin{lemma}\label{wplsdord}
Let $H$ be a semicomplete digraph with possible loops. Suppose that
$L$ does not contain either $R$ or $\vec{C}^*_3$ as an induced
subdigraph, and $U(L^{sym})$ is a proper interval graph. Also
suppose that $I$ is a transitive tournament and $H$ does not contain
either $W$, $R'$ or $\vec{C}_3$ with at least one loop as an induced
subdigraph. Then the vertices of $H$ can be ordered $v_1,\ldots
,v_n$ so that for every pair of vertices $v_i$ and $v_j$ with $i<j$,
we have $v_i\rightarrow v_j$.
\end{lemma}
\pf Let $L^{sym}_1, \ldots ,L^{sym}_l$ be the acyclic ordering of
the components of $L^{sym}$. Let $$w_1,w_2,\ldots
,w_q=v_1^1,v^1_2,\ldots ,v_{|V(L^{sym}_1)|}^1,\ldots ,v_1^i,v_2^i,
\ldots ,v_{|V(L^{sym}_i)|}^i,\ldots , v_{1}^l,v_1^l,\ldots
,v_{|V(L^{sym}_l)|}^l$$ be the ordering of $V(L)$ as described in
Lemma \ref{refsdord}. Let $u_1, \ldots , u_p$ be the acyclic
ordering of $V(I)$, i.e., $u_i\dom u_j$ implies $i<j.$ We will prove
the statement by showing that the subdigraph induced by
$V(L^{sym}_i)$ can be 'inserted' into an appropriate position among
the acyclic ordering of $V(I)$ without creating a cycle, thus by
constructing an ordering of $V(H)$ satisfying the asserted property.

Observe that any arc between a pair of two vertices from $I$ and $L$
is asymmetric. Otherwise, the two vertices induce a digraph $W$,
which is impossible by the assumption.

First, we claim that given a vertex $u$ of $I$ and a component
$L^{sym}_i$, we have either $u \mapsto V(L^{sym}_i)$ or
$V(L^{sym}_i) \mapsto u$. If $L^{sym}_i$ is a trivial component
consisting of a single vertex, the claim follows directly. So,
assume that $|V(L^{sym}_i)| \geq 2$ and there exists two vertices
$v$, $v'$ of $L^{sym}_i$ such that $u \mapsto v$ and $v' \mapsto u$.
Then, since $L^{sym}_i$ is connected, there is a path in $L^{sym}_i$
linking $v$ and $v'$. We can find two adjacent vertices $s,t$ on
this path such that $u\mapsto s$ and $t \mapsto u$. However, then
$H[\{u,s,t\}]\cong R'$, a contradiction.

Secondly, we claim that for each component $L^{sym}_i$ (possibly
trivial), the vertices of $L^{sym}_i$ can be `inserted' into an
appropriate position so that the ordering of $V(I) \cup
V(L^{sym}_i)$ satisfies the required property. (Here, the ordering
within $V(L^{sym}_i)$ remains unchanged.) That is, either $V(I)
\mapsto V(L^{sym}_i)$ or $V(L^{sym}_i) \mapsto V(I)$, or there exist
an integer $1 \leq j < p$ such that for all $k \leq j$, we have $u_k
\mapsto V(L^{sym}_i)$ and for all $k > j$, we have $V(L^{sym}_i)
\mapsto u_k$. If $V(I) \mapsto V(L^{sym}_i)$ or $V(L^{sym}_i)
\mapsto V(I)$, the ordering $u_1, \ldots , u_p$ followed by or
following the ordering of $V(L^{sym}_i)$ trivially satisfies the
required property. Thus, we may assume that $u \mapsto V(L^{sym}_i)$
and $V(L^{sym}_i) \mapsto u'$ for some $u,u' \in V(I)$.

Suppose that there are two vertices $u_j$ and $u_{j'}$ of $I$ with
$j' < j$ such that $u_j \mapsto V(L^{sym}_i)$ and $V(L^{sym}_i)
\mapsto u_{j'}$. Then $u_{j'}, u_j$ together with a vertex of
$L^{sym}_i$ form $\vec{C}_3$ with a loop, contradicting the
assumption. Hence, if $u_j \mapsto V(L^{sym}_i)$ for some $u_j \in
V(I)$, then $u_{j'} \mapsto V(L^{sym}_i)$ for each $j' < j$.
Similarly, if $V(L^{sym}_i) \mapsto u_j$ for some $u_j \in V(I)$,
then $V(L^{sym}_i) \mapsto u_{j'}$ for each $j'
> j$. By taking the maximum $j$ such that $u_j \mapsto V(L^{sym}_i)$

and inserting $V(L^{sym}_i)$ between $u_j$ and $u_{j+1}$ while
preserving the ordering within $V(L^{sym}_i)$, we are done with the
claim. From now on, we will say that {\em $V(L^{sym}_i)$ is inserted
after $u_j$} if $u_j \mapsto V(L^{sym}_i)$ and $V(L^{sym}_i) \mapsto
u_{j+1}$ when $u_{j+1}$ exists, and {\em $V(L^{sym}_i)$ is inserted
before $u_j$} if $V(L^{sym}_i) \mapsto u_j$ and $u_{j-1} \mapsto
V(L^{sym}_i)$ when $u_{j-1}$ exists.

Note that if any two components $L^{sym}_i$ and $L^{sym}_j$ of
$L^{sym}$ are inserted before/after the same vertex of $I$, we will
keep their relative order unchanged.

Now let us show that the insertion of all $V(L^{sym}_i)$'s does not
change their relative order in $L$. That is, if $V(L^{sym}_i)$ is
inserted after $u_j$, then each component $L^{sym}_{i'}$ for $i'>i$
is inserted after $u_{j'}$, where $j' \geq j$ and if $L^{sym}_i$ is
inserted before $u_j$, then each component $L^{sym}_{i'}$ for $i'<i$
is inserted before $u_{j'}$, where $j' \leq j$.

Suppose to the contrary that there are two components $L^{sym}_i$
and $L^{sym}_{i'}$ with $i <i'$ such that $V(L^{sym}_i)$ is inserted
after $u_j$ and $V(L^{sym}_{i'})$ is inserted before $u_{j'}$ with
$j' \leq j$. Then, by the above argument, $V(L^{sym}_{i'}) \mapsto
u_j$. However, a vertex from $V(L^{sym}_i)$, a vertex from
$V(L^{sym}_{i'})$ and $u_j$ induce $\vec{C}_3$ with two loops,
contradicting the assumption. Hence, if $V(L^{sym}_i)$ is inserted
after $u_j$, then $V(L^{sym}_{i'})$ for $i'>i$ is inserted after
$u_{j'}$, where $j' \geq j$. Similarly, we can show that if
$V(L^{sym}_i)$ is inserted before $u_j$, then $V(L^{sym}_{i'})$ for
$i'<i$ is inserted before $u_{j'}$, where $j' \leq j$.

It is straightforward from the above construction that the resulting
ordering satisfies the required property.\qed

Now we are ready to prove that $H$ has a Min-Max ordering when $H$
satisfies the condition (ii-a) in Theorem \ref{wplsd}.

\begin{lemma}\label{wplsdmm}
Let $H$ be a semicomplete digraph with possible loops. Suppose that
$L$ contains neither $R$ nor $\vec{C}^*_3$ as an induced subdigraph,
and  $U(L^{sym})$ is a proper interval graph. Also suppose that $I$
is a transitive tournament and $H$ does not contain either $W$, $R'$
or $\vec{C}_3$ with at least one loop as an induced subdigraph. Then
MinHOM($H$) has a Min-Max ordering.
\end{lemma}

\pf Consider an ordering $v_1, \ldots ,v_n$ of the vertices of $H$
as described in Lemma \ref{wplsdord}. We will show that this is a
Min-Max ordering of $V(H)$. Note that the induced ordering of $V(I)$
is an acyclic ordering and the induced ordering of $V(L)$ is a
Min-Max ordering for $L^{sym}$ as described in Lemma \ref{refsdord}.

Let $v_iv_j$ and $v_kv_l$ be any nontrivial pair of arcs of $H$.
Observe that if both arcs are in $A(L)$, then the minimum and the
maximum of them are also in $A(L)$ since the induced ordering of
$V(L)$ is a Min-Max ordering for $L$. Moreover, if both arcs are
forward arcs, i.e. $i<j$ and $k<l$, then we have either $i<k<l<j$ or
$k<i<j<l$. In either case, it follows from Lemma \ref{wplsdord} that
the minimum and the maximum of them are in $A(H)$.

Hence what we need to consider is the case where $v_kv_l$ is not a
forward arc. If $v_kv_l$ is a loop, then $i<k=l<j$. It follows from
Lemma \ref{wplsdord} that the minimum and the maximum of the two
arcs are in $A(H)$ in this case. Let $v_kv_l$ be a backward arc,
i.e., $k>l$. Clearly, $v_kv_l\in A(L)$. Then there are two remaining
cases to consider.

{\bf Case 1:} $v_iv_j\in A(I)$.

Then we have one of the following options: (a) $i<l<j<k$, (b)
$i<l<k<j$, (c) $l<i<k<j$, (d) $l<i<j<k$. However, in (a), $v_l
\mapsto v_j$ and $v_j \mapsto v_k$, which is a contradiction since
$v_k$, $v_l$ belong to the same component of $L^{sym}$, and $v_j$
has to either dominate or to be dominated by each component of
$L^{sym}$. With a similar argument, case (c) and case (d) are
impossible. By Lemma \ref{wplsdord}, in case (b), the minimum and
the maximum of $v_iv_j$ and $v_kv_l$ are in $A(H)$.

{\bf Case 2:} $v_iv_j \in A(H)\setminus (A(I)\cup A(L))$.

Since $v_iv_j \in A(H)\setminus (A(I)\cup A(L))$, exactly one of
$v_i$ and $v_j$ has a loop. Assume that $v_j$ has a loop. The case
for which $v_i$ has a loop can be treated in a similar manner.

Then we have one of the following options: (a) $i<l<j \leq k$, (b)
$i<l<k<j$, (c) $l<i<k \leq j$, (d) $l<i<j<k$. However, if (c) is the
case, $v_l \rightarrow v_i$ and $v_i \rightarrow v_k$, which is a
contradiction since $v_k$, $v_l$ belong to the same component of
$L^{sym}$ and $v_i$ has to either dominate or be dominated by each
component of $L^{sym}$. With a similar argument, case (d) is
impossible.

Let (a) be the case. Note that $v_k$ and $v_l$ belong to the same
component of $L^{sym}$. By the property of the ordering (see the
proof of Lemma \ref{wplsdord}), $v_j$ belongs to the same component
of $L^{sym}$ with $v_k$ and $v_l$. Since the ordering of the
vertices in this component satisfies the condition in Lemma
\ref{pincha}, $v_lv_k\in A(L^{sym})$ and $l<j \leq k$ imply that
$v_kv_j=v_{\max\{i,k\}}v_{\max\{j,l\}} \in A(L^{sym})$. By Lemma
\ref{wplsdord}, $v_iv_l=v_{\min\{i,k\}}v_{\min\{j,l\}} \in A(H)$.

Let (b) be the case. By Lemma \ref{wplsdord}, both the minimum and
the maximum of the two arcs are in $A(H)$. \qed

\begin{theorem}\label{wplsdpoly}
Let $H$ be a semicomplete digraph with possible loops. If one of the
followings holds, then MinHOM($H$) is polynomial time solvable.

(a) The digraph $H=\vec{C}_k$ for $k=$2 or 3.

(b) The digraph $L$ does not contains either $R$ or $\vec{C}^*_3$ as
an induced subdigraph, and $U(L^{sym})$ is a proper interval graph;
$I$ is a transitive tournament; $H$ does not contain either $W$,
$R'$ or $\vec{C}_3$ with at least one loop as an induced subdigraph.

\end{theorem}
\pf Consider the following cases.

{\bf Case 1:} The condition (a) holds. Then, there is a polynomial
time algorithm for MinHOM($H$). We give the algorithm for the sake
of completeness. We consider $H=\vec{C}_k$ with an arbitrary integer
$k \geq 2$. We assume that the input digraph $D$ is connected since
otherwise, the algorithm can be applied to each component of $D$ and
we can sum up the costs of homomorphisms of each component to $H$.

Choose a vertex $x$ of $D$, and assign it color 1. For any vertex
$y$ with color $i$, we assign all the in-neighbors of $y$ color
$i-1$ and all the out-neighbors of $y$ color $i+1$, where the
operation is taken modulo $k$. It is easy to see that no vertex of
$D$ is assigned a pair of conflicting colors if and only if $D$ has
a $\vec{C}_k$-coloring. Furthermore, cyclicly permutating the colors
of $V(D)$ does not affect the existence of a homomorphism of $D$ to
$H$. Hence, we can assign $x$ color $2,\ldots ,k$, modify the
assignment of other vertices of $D$ accordingly, and compute the
cost of homomorphism respectively. We finally accept an assignment
which leads to the minimum cost.

{\bf Case 2:} The condition (b) holds. Then by Lemma \ref{wplsdmm}
and Theorem \ref{mmth}, MinHOM($H$) is polynomial time solvable.
\qed

\begin{corollary}
Let $H$ be a semicomplete digraph w.p.l. Then MinHOM($H$) is
polynomial time solvable if $H=\vec{C}_k$ for $k=$2 or 3, or $H$ has
a Min-Max ordering. Otherwise, MinHOM($H$) is NP-hard.
\end{corollary}

\subsection{Proving that (ii-a) and (ii-b) of Theorem \ref{wplsd} are
equivalent}\label{wplchar}

In this subsection, we will prove that (ii-a) and (ii-b) in Theorem
\ref{wplsd} are equivalent.

It follows from the proof of Lemma \ref{wplsdord} that the condition
(ii-a) implies (ii-b). Indeed, from the construction of the ordering
in the proof of Lemma \ref{wplsdord}, $H$ is a composition digraph,
i.e., $H=TT_{p+l}[S_1,S_2,\ldots ,S_{p+l}]$ where $S_i$ for each
$i=1, \ldots ,p+l$ is one of the two types: (a) a single vertex
without a loop, (b) a reflexive semicomplete digraph which does not
contain $R$ as an induced subdigraph, and for which $U(S^{sym}_i)$
is a connected proper interval graph. Here, $p$ is the number of
vertices in $V(I)$ and $l$ is the number of components (possibly
trivial) of $L^{sym}$.

Lemma \ref{eqv} given below shows that the converse is also true,
accomplishing the equivalence of (ii-a) and (ii-b) in Theorem
\ref{wplsd}.

For further reference, we give a well-known theorem that
characterizes proper interval graphs in terms of forbidden
subgraphs. We will start with some definitions. A graph $G$ is
called {\em a claw } if $V(G)=\{x_1,x_2,x_3,y\}$ and
$E(G)=\{x_1y,x_2y,x_3y\}.$ A graph $G$ with
$V(G)=\{x_1,x_2,x_3,y_1,y_2,y_3\}$ is called {\em a net} if
$E(G)=\{x_1x_2,x_2x_3,x_3x_1,y_1x_1,y_2x_2,y_3x_3\}$, and {\em a
tent } if
$E(G)=\{x_1x_2,x_2x_3,x_3x_1,y_1x_2,y_1x_3,y_2x_1,y_2x_3,y_3x_1,y_3x_2\}.$

\begin{theorem}\cite{hellSIDMA18}\label{pinfbd}
A graph $G$ is a proper interval graph if and only if it does not
contain a cycle of length at least four, a claw, a net, or a tent as
an induced subgraph.
\end{theorem}

\begin{lemma}\label{eqv}
Let $H=TT_{k}[S_1,S_2,\ldots ,S_{k}]$ where $S_i$ for each $i=1,
\ldots ,k$ is either a single vertex without a loop, or a reflexive
semicomplete digraph which does not contain $R$ as an induced
subdigraph and for which $U(S^{sym}_i)$ is a connected proper
interval graph. Then, $H$ is a semicomplete digraph w.p.l. such that
$L$ does not contain either $R$ or $\vec{C}^*_3$ as an induced
subdigraph, and $U(L^{sym})$ is a proper interval graph, $I$ is a
transitive tournament and $H$ does not contain either $W$, $R'$ or
$\vec{C}_3$ with at least one loop as an induced subdigraph.
\end{lemma}
\pf Clearly, $H$ is a semicomplete digraph w.p.l. and $I$ is a
transitive tournament. Furthermore, the absence of $W$, $R'$ or
$\vec{C}_3$ with one or two loops in $H$ follows from the transitive
tournament structure of $H$.

Therefore, it remains to show that $L$ does not contain
$\vec{C}^*_3$ as an induced subdigraph. To the contrary, suppose
that there are vertices $u,v,w\in V(L)$ such that $H[\{u,v,w\}]\cong
\vec{C}^*_3$ ($u\mapsto v\mapsto w\mapsto u$). Then $u,v$ and $w$
must belong to the same component $S_i$. Since $S_i^{sym}$ is
connected, there exist paths between any pair of vertices in
$\{u,v,w\}$. Define
$\mu(u,v,w)=\min\{\dist(u,v),\dist(u,w),\dist(v,w)\}$, where
$\dist(x,y)$ is the length of a shortest path between $x$ and $y$ in
$S_i^{sym}$.

Choose a triple $u,v,w$ in $S_i$ such that $H[\{u,v,w\}]\cong
\vec{C}^*_3$ ($u\mapsto v\mapsto w\mapsto u$) and $\mu(u,v,w)$ is
minimal. Assume that $\dist(u,v)=\mu(u,v,w).$ Consider a shortest a
path $P=u(=u_0),u_1,\ldots ,u_p(=v)$ between $u$ and $v$ in
$S_i^{sym}$. Observe that $\dist(u,v)\ge 2.$ Let $a_v$ ($a_w$) be an
arc between $v$ and $u_1$ (between $w$ and $u_1$). If both $a_v$ and
$a_w$ are symmetric, then $u,v,w$ and $u_1$ form a claw in
$S_i^{sym}$, which is impossible by Theorem \ref{pinfbd}. Hence, at
most one of $a_v$ and $a_w$ is symmetric.

If $a_v$ is symmetric, then $a_w$ must be asymmetric and we have
either $H[\{u,w,u_1\}]\cong R$ or $H[\{v,w,u_1\}]\cong R$, a
contradiction. Similarly, if $a_w$ is symmetric, we have either
$H[\{u,v,u_1\}]\cong R$ or $H[\{w,v,u_1\}]\cong R$, also a
contradiction. Hence, both $a_v$ and $a_w$ are asymmetric. Suppose
that $u_1\mapsto w$. Then $H[\{u,u_1,w\}]\cong \vec{C}^*_3$, a
contradiction. Hence, $w\mapsto u_1$. Similarly, $u_1\mapsto v.$
Thus, $u_1,v,w$ is a triple with $H[\{u_1,v,w\}]\cong \vec{C}^*_3$
such that $\mu(u_1,v,w)<\mu(u,v,w)$, a contradiction to the choice
of $u,v,w$.

Thus, $L$ does not contain $\vec{C}_3^*$ as an induced subdigraph,
which completes the proof. \qed

\section{Further Research}

We obtained a dichotomy classification for reflexive semicomplete
digraphs and semicomplete digraphs w.p.l. This solves the question
raised in our previous paper \cite{gutinRMS}. The obtained results
imply that given a (loopless) semicomplete digraph $H$, for
MinHOM($H$) to be polynomial time solvable, $H$ should be a very
simple directed cycle or it has to be acyclic. On the other hand, in
the dichotomy for reflexive semicomplte digraphs, Min-Max ordering
appears to play the central role to characterize the polynomial time
solvable cases. We'd like to conjecture that for a reflexive digraph
$H$, MinHOM($H$) is polynomial time solvable if and only if $H$ has
a Min-Max ordering. The comparison between this conjecture and the
proper ordering conjecture suggested by Feder et. al. in
\cite{federMAN} for ListHOM($H$), when $H$ is a reflexive digraph,
presents an interesting point of view.

The problem of obtaining a dichotomy classification for semicomplete
$k$-partite digraphs, $k\ge 2$, w.p.l. remains still open. In fact,
even settling a dichotomy for $k$-partite tournaments  seems to be
not easy. Actually, for a $k$-partite tournament w.p.l. $H$, a
complete dichotomy of MinHOM($H$) has been obtained in
\cite{gutinRMS} provided that $H$ has a cycle.

\end{document}